\documentstyle[twoside,fleqn,espcrc2,epsfig]{article}

\newcommand{\AmS}{{\protect\the\textfont2
  A\kern-.1667em\lower.5ex\hbox{M}\kern-.125emS}}
\newcommand{\be}{\begin{equation}}
\newcommand{\eea}{\end{eqnarray}}
\newcommand{\beq}{\begin{equation}}
\newcommand{\eeq}{\end{equation}}
\newcommand{\bea}{\begin{eqnarray}}
\newcommand{\ee}{\end{equation}}

\def\Journal#1#2#3#4{{#1} {\bf #2} (#4) #3}

\def\NPB{\em Nucl. Phys.}
\def\PLB{\em Phys. Lett.}

\def\CMP{\em Commun. Math. Phys.}
\title{\vspace*{-0.1cm}
MATRIX ELEMENTS OF LIGHT QUARKS}
\author{G.~Martinelli\address{\vspace*{-0.28cm} Dip. di Fisica, Univ. ``La Sapienza"  and 
INFN, Sezione di Roma, P.le A. Moro, I-00185 Rome, Italy.}}
\begin{document}
\newcommand{\sze}{\small}
\begin{abstract}
\vspace*{-0.1cm}
Lattice results for matrix  elements of light quark operators  are reviewed.  The discussion is focused on recent 
theoretical progress and new numerical calculations  which appeared after 
the Lattice 2000 Conference.
\vspace*{-0.6cm}
\end{abstract}
\maketitle
\section{Introduction}
\vspace{-0.1cm} 
\label{sec:intro}
In this talk  a review  of theoretical and numerical results  for matrix elements of light quark operators is presented.    
The review  is focused on   the definition, renormalization  and improvement of operators relevant to 
different processes and on the  evaluation of the corresponding physical amplitudes. 
At this Conference the main topics  covered 
in the parallel  sessions  were  electromagnetic and weak form factors, 
structure functions, $K^{0}$--$\bar K^{0}$ mixing, kaon decay  amplitudes and
light-cone wave functions. Correspondingly the problem of the renormalization and improvement of the operators was also discussed. 
Results with  perturbative and non-perturbative, gauge invariant or gauge dependent  renormalization methods, using Ward 
identities, with unimproved or improved operators, with Wilson like  (improved, twisted, \ldots), 
Domain Wall or overlap fermions   were presented.  In total 
I counted 38 talks in the parallel sessions and 13 posters. It is 
clear that it would have been impossible to discuss all this material, 
to which two accurate and complete review 
talks  by L.~Lellouch~\cite{lellouch2000} and S.~Sint~\cite{sint2000} 
were dedicated  at Bangalore. A selection 
of the subjects  has been then unavoidable and I apologize to the authors 
of submitted contributions and to the speakers
of talks given  in the parallel sessions which I had to omit in my 
review. I also warmly  thank all  collegues who have kindly sent material and information for the preparation of this talk.

The most exciting news at this Conference  are the intriguing results for $\epsilon^{\prime}/\epsilon$ presented by 
the CP-PACS~\cite{cppacs2001} and RBC~\cite{rbc2001} Collaborations. 
Thus I  start from these results to discuss a few  physics issues of general interest.  The renormalization of lattice quark 
operators, the {\it ultraviolet}  problem, 
 and the calculation of final state interaction effects, the {\it infrared} problem,  
which is 
essential for kaon decays, are  the next subjects.  This will give me 
the opportunity to show several  results that were presented in parallel talks at this Conference.  More physics problems which are 
interesting for the lattice comunity can be found in the talk by  M.~Beneke at this Conference~\cite{benekelatt2001} and by J.~Donoghue 
at the Kaon 2001 Conference~\cite{donoghuekaon2001}.  

The main topics  in this review  are:  i) the determination of  the $\Delta 
I=1/2$ and $\Delta  I=3/2$ $K \to  \pi\pi$ 
amplitudes~\cite{cppacs2001,rbc2001,dlin2001,spqr};  ii) the main contributions to  $\epsilon^{\prime}/\epsilon$ due to the electropenguin 
operators ($Q_{7}$ and $Q_{8}$ in the standard  notation) and to the strong 
penguin operator $Q_{6}$~\cite{cppacs2001,rbc2001,dlin2001,spqr};  
iii)  matrix elements of SUSY operators which can be  computed on the 
lattice~\cite{damir2}; iv)  studies of the chiral expansion of 
the   relevant amplitudes  in finite and infinite 
volumes, in the  quenched and unquenched 
cases~\cite{bernardgolterman}--\cite{goltepallalast}; v) studies of Final State 
Interactions (FSI) in lattice simulations~\cite{spqr}; vi) the extraction of the 
physical amplitudes from matrix elements computed in the Euclidean   space-time on a finite volume~\cite{lll,lmst}. 
The problem of the renormalization and improvement of the relevant operators 
will be mentioned and discussed along with the presentation of  the different 
arguments. Although the discussion is mainly focused on kaon physics, many 
considerations are general and apply to  the calculation of all 
matrix elements.
\vspace{-0.2cm}\section{General physics issues}\label{sec:general}
\vspace{-0.1cm} 
\begin{table*}[htb]
\caption{{\sl \small Experimental measurements and theoretical predictions 
for several neutral current processes: the CP violation parameters 
$\epsilon$ and $\epsilon^{\prime}/\epsilon$, the neutral $B$ meson 
mixing $\Delta M_{s}$ and $\Delta M_{d}$,  radiative  inclusive $b 
\to s\gamma$ decays and $K^{+} \to \pi^{+} \nu \bar \nu $. In the 
latter two cases the calculation can be done by using the HQET 
expansion and the experimental $K \to \pi$ weak current matrix 
element respectively. Thus  a lattice calculation of the hadronic amplitudes is not needed.}}
\label{tab:fcnc}
\newcommand{\m}{\hphantom{$-$}}
\newcommand{\cc}[1]{\multicolumn{1}{c}{#1}}
\renewcommand{\tabcolsep}{2pc} 
\begin{tabular}{@{}lll}
\hline
Quantity& Experiments &Theory \\
\hline
$\epsilon$ \ & $2.271 \pm 0.017 \times 10^{-3}$~\cite{PDG} & $\eta \, (1-\rho) \,  B_{K} $  \\
$\epsilon^{\prime}/\epsilon$ & $17.2 \pm 1.8\times 10^{-4}$ 
~\cite{na48,ktev}   & $-7 \div 30 \times 10^{-4}$ \\
$\Delta M_{s} /\Delta M_{d}$ &  $> 30  \,  (95\% C.L.)$ &  $[(1-\rho)^{2}+\eta^{2}¥]^{-1} \xi$ \\
$BR(B \to X_{s} \gamma)$ & $ 3.11 \pm 0.39 \times 
10^{-4}$~\cite{bsgexp} & $ 3.50 \pm 
0.50 \times 10^{-4}$ No Lattice Needed \\ 
$BR(K^{+} \to \pi^{+} \nu \bar \nu )$ & $1.5^{+3.4}_{-1.2} \times 
10^{-10}$~\cite{kpnnexp} & $ 0.8\pm 
0.3 \times 10^{-10}$ No Lattice Needed \\ 
\hline
\end{tabular}\end{table*}
\begin{table*}[htb]
\caption{{\sl \small Lattice results for $\Delta I=1/2$ transitions using $K \to \pi$ 
matrix elements from RBC and  CP-PACS. The experimental numbers are 
also given.}}
\label{tab:rbcpacs}
\newcommand{\m}{\hphantom{$-$}}
\newcommand{\cc}[1]{\multicolumn{1}{c}{#1}}
\begin{tabular}{@{}lllll}
\hline
Reference & $Re {\cal A}_{0}$ & $Re {\cal A}_{2}$ & $Re {\cal A}_{0}/Re {\cal A}_{2}$   & $\epsilon^{\prime} / \epsilon$ \\
\hline \hline
CP-PACS~\cite{cppacs2001} & $16 \div 21 \times 10^{-8}$ & $1.3 \div 1.5 \times 10^{-8}$ & 
$9 \div 12$ &  $-7 \div -2 \times 10^{-4}$\\
RBC~\cite{rbc2001}  & $29 \div 31 \times 10^{-8}$ & $1.1 \div 1.2 \times 
10^{-8}$ & $24 \div 27$  &  $-8 \div -4 \times 10^{-4}$ \\
Exps.~\cite{PDG,na48,ktev} 2001& $33.3 \times 10^{-8}$ &    $1.5 \times 
10^{-8}$ &  $ 22.2 $ & $17.2 \pm 1.8  \times 10^{-4}$ \\
\hline
\end{tabular}\\[2pt]
\end{table*}
Flavour changing neutral currents (FCNC) have been  and are the main source  
of  information on quark masses, couplings and CP violation. It is sufficient to 
mention that they led to the GIM theory and that the large $B$--$\bar 
B$ mixing was the first  strong indication that the top quark was very 
heavy, long before  precision tests at LEP or its discovery. They are  
also a window on physics beyond the Standard Model (SM). Since in the 
SM  FCNC cannot occur at tree level, their effects  compete with contributions 
from other heavy sectors of the theory as the extra particles predicted, for example, by 
Supersymmetry. This explains the large experimental and theoretical 
efforts in FCNC studies. In table~\ref{tab:fcnc}, a list of FCNC 
processes for which experimental measurements exist is presented. In 
two cases the comparison of the theory with experiments requires the 
knowledge of the hadronic matrix elements. Lattice results for 
the matrix element of the $\Delta S=2$ operator
relevant for  $\epsilon$, encoded in the bag  parameter $\hat B_{K}$,  
exist since a long time.  This year, for the first time, we  have two complete 
calculations of the matrix elements  of the operators entering  $\epsilon^{\prime} / \epsilon$
 (and also  $Re {\cal A}_{0}$ and 
$Re {\cal A}_{2}$) which can be compared with the experimental 
results. Here I want first to  examine the  
physics  implications of the RBC and CP-PACS calculations   which are in 
total disagreement with the experimental measurements of $\epsilon^{\prime} / 
\epsilon$ by NA48 and KTeV, table~\ref{tab:rbcpacs}. If 
confirmed, these results  imply other contributions to  $\epsilon^{\prime} / 
\epsilon$ from physics beyond the SM.  Let me give an explicit example 
which could explain the discrepancy.  In the 
effective Hamiltonian, besides the dimension-six four fermion 
operators considered in refs.~\cite{cppacs2001,rbc2001}, there are two
dimension-five  chromomagnetic operators of the form
\vspace*{-0.2cm} \be {\cal H}_{g} = C^{+}_{g} Q^{+}_{g} + C^{-}_{g} Q^{-}_{g}  \, , \ee
\vspace*{-0.2cm}
where
\vspace*{-0.2cm}\be Q^{\pm}_{g}= \frac{g}{16\pi^{2}} \left(\bar s_{L} \sigma^{\mu\nu} 
t^{a} d_{R}  G^{a}_{\mu\nu} \pm \bar s_{R} \sigma^{\mu\nu} 
t^{a} d_{L}  G^{a}_{\mu\nu} \right) \, . \ee \vspace*{-0.2 cm}
The chromomagnetic contribution  exists  in the SM (although it has 
not been evaluated by~\cite{cppacs2001,rbc2001})  but it is expected 
to give a rather small contribution to $\epsilon^{\prime} / 
\epsilon$  since it is chirally suppressed ($\sim 
{\cal O}(m_{K}^{4})$)~\cite{bertolini}. SUSY effects may enhance the  
chromomagnetic  term~\cite{murayama} and this could be the explanation of the 
difference between  lattice results and  experiments.  A
large  chromomagnetic term may  trigger visible effects  in 
hyperon decays~\cite{murayama2} and in CP violating 
$K \to \pi\pi\pi$ amplitudes~\cite{isidorid}. Its coupling is 
constrained also by $\epsilon$ since it may  give a large contribution 
there, although this is rather hard to estimate. Thus for the lattice comunity it is certainly very interesting 
to compute  the relevant amplitudes of the chromomagnetic 
operators, similarly to what has been done for the electromagnetic 
one, which enters  the CP violating $K_{L} \to 
\pi^{0}e^{+}e^{-}$ decay~\cite{damir1}.

FCNC exist for different changes of flavour: 
\be {\cal H}^{FCNC} =  {\cal H}^{\Delta F=0} +{\cal 
H}^{\Delta F=1} +{\cal H}^{\Delta F=2} \, .\ee
$\Delta F=1$, corresponding to  $\epsilon^{\prime} / 
\epsilon$, and   $\Delta F=2$ processes, corresponding to  $\epsilon$ 
and $B$--$\bar B$ mixing, are the 
most studied.  Although they  are of fundamental 
importance for the strong CP problem,  $\Delta F=0$ FCNC processes have not received, instead, the 
necessary attention by the lattice community. $\Delta F=0$ FCNC  
processes   are also important because from them one may derive very
stringent  constraints for any   extension of the SM.  In this 
framework  particular important is the electric dipole moment of the 
neutron for which a very tight upper bound exists, $d_{{\cal N}}< 6.3  
\times 10^{-26}$ e cm~\cite{dnkaon2001}. After the first studies at the 
end of the 80s~\cite{agms}, the calculation of the neutron electric dipole moment 
has been  abandoned. This is very surprising since
the renormalization of the relevant  operators 
($G^{a}_{\mu\nu} \tilde G^{a}_{\mu\nu}$ ot $m \bar q \gamma_{5} q$)  and 
the calculation of the necessary disconnected
diagrams with stochastic sources is by now a common practice.
There are several extra operators contributing to $d_{{\cal N}}$ that appear in extension of the SM and 
the matrix elements of which have never been computed~\cite{carlos}
\bea  {\cal H}_{eff}^{\Delta F=0} &=& -\frac{i}{2} C_{e} \, \bar q \sigma^{\mu\nu} 
\gamma_{5} q  F_{\mu\nu} -\frac{i}{2} C_{C} \, \bar q \sigma^{\mu\nu} 
\gamma_{5} t^{a} q G^{a}_{\mu\nu}\nonumber \\ &-&\frac{1}{6}  C_{g} \,  f_{abc} 
G^{a}_{\mu\rho} G^{b\rho}_{\nu} G^{c}_{\lambda\sigma} 
\epsilon_{\mu\nu\lambda\sigma}\, ,  \eea
where $C_{e,C,g}$ are the model dependent Wilson coefficients of the 
effective low energy theory.
After this brief discussion of FCNC amplitudes which could be computed 
on the lattice with interesting implications, let us return to the 
results which have been presented at this Conference and to the 
progress which have been achieved since  Lattice 2000. 
\vspace{-0.2cm}
\section{General framework}
\vspace{-0.1cm} 
Physical kaon weak decay amplitudes can be described with 
negligible error  (of ${\cal O}(\mu^{2}/M_{W}^{2})$, where $\mu$ is 
the renormalization scale) in terms of   matrix elements of the effective 
weak Hamiltonian 
$ \langle \pi \pi \vert {\cal H}_{eff}^{\Delta S=1} \vert K 
\rangle$,
 written as combination of  
Wilson coefficients and renormalized local  operators 
\beq {\cal H}_{eff}^{\Delta S=1}= -\frac{G_{F}}{\sqrt{2}} \sum_{i}  C_{i}(\mu) 
\hat Q_{i}(\mu) \,  .  \eeq 
The sum is over a complete set of operators, which depend on  $\mu$. In general there are 12 four-fermion operators 
and  two  dimension-five  operators: a  chromomagnetic one and an 
electromagnetic one. As discussed before, in the SM the contribution of the dimension-five operators 
is usually neglected.   The calculation of the matrix elements 
must be done non-perturbatively, and this is the r\^ole of the 
lattice, whereas the Wilson coefficients can be computed 
in perturbation theory.
\vspace{-0.2cm}\subsection*{Wilson coefficients and renormalized operators}\vspace{-0.1cm}
For ${\cal H}_{eff}^{\Delta S=1}$  the Wilson 
coefficients have been computed at the next-to-leading order~\cite{Buras:2001tc}.
 The perturbative calculation  is reliable  provided that the scale $\mu$ is large enough, $\mu 
\gg \Lambda_{QCD}$. In this 
respect calculations performed below the charm quark mass ($m_{c} \sim 
1.3$~GeV)  are, in my opinion, suspicious. In fact, either $\mu \ll 
m_{c}$, and then perturbation theory is questionable, or $\mu \sim 
m_{c}$, and then the effective three-flavour weak Hamiltonian (with 
propagating up, down and strange quarks) cannot 
be properly matched to the four-flavour theory (up,down, strange and charm) 
because of the presence of  operators of higher dimension which 
contribute at ${\cal O}(\mu^{2}/m_{c}^{2})$.  
  Wilson coefficients and   matrix elements of the  operators 
$\hat Q_{i}(\mu)$ separately
depend on the choice of the renormalization scale and scheme. In the lattice approach,
it is possible to show that  this dependence  cancel in
physical quantities, up to higher-order corrections in the perturbative expansion of
the Wilson coefficients. 

Matching of  bare (divergent) lattice 
operators, $Q_{i}(a)$ to the renormalized ones is obtained  by computing suitable 
renormalization constants $Z_{ik}(\mu a)$
\bea {\cal A}^{i}_{I=0,2}(\mu) &=& \langle \pi \pi \vert \hat Q_{i}(\mu)  \vert K
\rangle_{I=0,2}\nonumber \\ &=& \sum_{k} \, Z_{ik}(\mu a) \langle \pi \pi \vert  Q_{k}(a)  \vert K
\rangle_{I=0,2} \, , \label{eq:renor}\eea
where $a$ is the lattice spacing.
The  \textit{ultra-violet} (UV) problem,
which deals with the construction of finite matrix elements of
renormalized operators constructed from the bare lattice ones, has
been addressed in a series of papers~\cite{wise}-\cite{capitani} 
and is, at least in principle, completely solved. The remaining 
difficulties are practical ones. On the one hand, the $\Delta I=1/2$ 
operators suffer from  power divergences in the ultraviolet cutoff, 
$1/a$.  These divergences, that cannot be subtracted using  perturbation 
theory~\cite{lellouch2000},  can be eliminated by performing suitable 
numerical  subtractions. The subtraction procedure, however, suffers from systematic uncertainties which are 
difficult to keep under control,  see below.  On the other hand, the  
perturbative calculation of   the logarithmically divergent (or 
finite)    $Z_{ik}(\mu a)$   is rather inaccurate. Several non-perturbative methods have been 
developed in order to compute $Z_{ik}(\mu a)$   
non-perturbatively~\cite{npm}\--\cite{lleshouches} and the uncertainties 
vary between  1\%, for the simplest bilinear operators, to $10\div 25$\%, in the case of the 
four-fermion operators of interest. An accurate determination of 
$Z_{ik}(\mu a)$  for the $\Delta I = 1/2$ operators  has not been 
achieved to date  and more work is needed in this direction.  Let me 
also recall a method for a gauge-invariant  non-perturbative renormalization of 
the relevant operators which was proposed in~\cite{rossi} and never 
tried, in spite of the fact that it should be rather easy to implement. 
This proposal is based on the study, at short Euclidean distances 
$\vert x \vert \ll 1/\Lambda_{QCD}$, of suitable correlators. For example, in order to 
renormalize bilinear operators of the form $J(x) = \bar q(x) \Gamma 
q(x)$,  one should compute numerically $\langle 0 \vert 
J(x)^{\dagger} J(0)\vert 0\rangle$ and compare the results with the 
expression obtained in perturbation theory in a given popular 
renormalization scheme ($\overline{MS}$ or $RI$-$MOM$ for example).  The 
method  avoids the problem of the window,  $a \ll \vert x \vert \ll 
1/\Lambda_{QCD}$,  by successive matching of the renormalization 
conditions~\cite{lleshouches} and can be  extended to four 
fermion operators. For bilinear operators, the necessary two-loop 
perturbative calculations, including the finite terms, have been 
completed and the extension to four fermion operators is 
underway~\cite{delbello}. 

Since numerical simulations 
are performed at finite values of the  lattice spacing, $a^{-1} \sim 
2\div 4$~GeV, another source of uncertainty in the determination of the matrix 
elements  comes from discretization errors. They  are of ${\cal O}(\Lambda_{QCD} a)$,  ${\cal  O}(\vert \vec p\vert a)$ or ${\cal  
O}(m_{q} a)$, where $\vec p$ is a typical  hadron momentum and $m_{q}$ the quark mass.  The  simplest strategy to reduce discretization effects  
consists in   computing physical  quantities at several values of the lattice spacing and then 
extrapolate to $a \to 0$.  A different  approach is to reduce discretization errors from 
${\cal O}(a)$ to ${\cal O}(a^{2})$ by improving the lattice action 
and operators~\cite{lleshouches}. A systematic study of the improvement of four fermion 
operators is still to be done. Note that with DWF~\cite{dwf} or overlap 
Fermions~\cite{of} the errors are automatically of  ${\cal 
O}(a^{2})$~\cite{pilar}.   Discretization errors correspond to the 
matching problems in effective theories with a low cutoff recently 
discussed  by Cirigliano, Donoghue and Golowich~\cite{cdg}. In this 
respect the problem is softer for the lattice approach since numerical 
simulations are already performed at relatively large scales.
\vspace{-0.25cm}\subsection*{$K\to \pi$ and $K \to \pi\pi$ matrix elements}\vspace{-0.1cm}
Two main roads have been suggested in the past in order to obtain ${\cal 
A}^{i}_{I=0,2}(\mu)$:

i) Compute the $K \to 0$ and $K \to \pi$ matrix elements $\langle 
0 \vert \hat Q_{i}(\mu)  \vert K \rangle$ and $\langle \pi \vert \hat Q_{i}(\mu)  \vert K
\rangle$  and then derive $\langle \pi \pi \vert \hat Q_{i}(\mu)  \vert K
\rangle_{I=0,2}$  using soft pion 
theorems~\cite{wise,renorm}. In this case the $K \to 
\pi\pi$ amplitudes can be evaluated only at the lowest order of the 
chiral expansion and FSI effects in the physical amplitudes are lost. 
 This  problem  poses serious difficulties  to lattice calculations based on the 
extraction of the $K \to \pi\pi$ amplitudes from the $K \to \pi$ 
matrix elements using chiral perturbation theory ($\chi$PT), as done by the 
CP-PACS~\cite{cppacs2001} and RBC Collaborations~\cite{rbc2001}.
The importance of FSI effects  for  $\Delta I=1/2$ transitions was first noticed in 
refs.~\cite{truong,isgur}. More recently it has  been  emphasized  by   Bertolini, 
Eeg and Fabbrichesi~\cite{bertolini} and  Pallante and 
Pich~\cite{PP} that they may have large effects also for $\epsilon^{\prime}/\epsilon$.
 Although the quantitative evaluation of  FSI effects  is still 
controversial~\cite{burasrome}, there is a general 
consensus that,  for a reliable calculation of kaon decay amplitudes,
it is necessary to have a good theoretical control of 
FSI. 

ii) Compute directly $\langle \pi \pi \vert \hat Q_{i}(\mu)  \vert K
\rangle_{I=0,2}$~\cite{lll,lmst,draper,dawson}.
The main difficulty in the latter case  is due to 
the relation between $K \to \pi\pi$ matrix elements computed in a
finite Euclidean space-time volume and the corresponding physical
amplitudes (the \textit{infrared} problem).  
The infrared problem arises from two sources:
i) the unavoidable continuation of the theory to
Euclidean space-time and
ii) the  use of a finite volume in numerical simulations.
An important progress has  been achieved by Lellouch and
L\"uscher (LL), who
derived a relation between the lattice $K\to\pi\pi$ matrix elements in a
finite volume and the physical kaon-decay amplitudes~\cite{lll}.
An alternative discussion of boundary
effects and the LL-formula, based on a study of the properties of
correlators of local operators was presented in ref.~\cite{lmst}. In this approach the LL-formula   
is derived for all elastic states under the inelastic threshold, with exponential
accuracy in the quantization volume. As a consequence  the relation between finite-volume matrix
elements and physical amplitudes, derived by LL for the lowest seven energy eigenstates, can be extended
to all elastic states under the inelastic threshold. It can also been
explicitly  demonstrated how finite volume correlators
converge to the corresponding ones in infinite volume.
The derivation of~\cite{lmst}  is based on the property of correlators of local
operators which can be expressed, with exponential accuracy, both
as a sum or as an integral over intermediate states, when
considering volumes larger than the interaction radius and
Euclidean times $0<t \simeq L$.  It can also been shown that it is possible to extract  $K \to \pi\pi$
amplitudes also when the kaon mass, $m_K$, does not match the
two-pion energy, namely when the inserted operator carries a
non-zero energy-momentum. Such amplitudes  are very useful, for
example, in the determination of the coefficients of  operators
appearing at higher orders in $\chi$PT, as illustrated by the numerical 
results for  $\Delta I=3/2$ transitions presented in sec.~\ref{sec:numerical}.

Let us sketch now the derivation  of the result with an illustrative 
example which is not explicitly written  in~\cite{lmst}.
Consider the following Euclidean T-products (correlators):
\bea G(t,t_{K}) &=& \langle 0 \vert T[J(t) \hat Q_{i}(0)   
K^{\dagger}(t_{K})]\vert  0\rangle \, ,  \nonumber \\
G(t) &=& \langle 0 \vert T[J(t) J(0)]   \vert  0\rangle \, ,\nonumber \\
G(t_{K}) &=& \langle 0 \vert T[ K(t_{K}) K^{\dagger}(0)\vert  0\rangle 
\, , 
\label{eq:corres} \eea
where $J$ is a scalar operator which excites (annhilates) zero angular 
momentum $\pi\pi$ states from (to) the vacuum and $K^{\dagger}$ is a pseudoscalar source which excites a kaon from the vacuum 
($t > 0$ and  $t_{K}  < 0$). At large time distances we have 
\bea   &&
G(t,t_{K}) \to \nonumber \\ && V \, \sum_{n} \langle 0 \vert J \vert  
\pi\pi(n)\rangle_{V} \langle  \pi\pi(n) \vert \hat Q_{i} \vert  K\rangle_{V}  \langle K \vert    K^{\dagger} \vert  0\rangle_{V} 
\nonumber  \\  && \exp(-W_{n} \, t -m_{K}\, t_{K}) 
\, ,  \eea  \bea
G(t) \to 
 V \, \sum_{n} \langle 0 \vert J \vert  
\pi\pi(n)\rangle_{V} \langle  \pi\pi(n) \vert J \vert 0\rangle_{V}  
\exp(-W_{n} \, t ) \nonumber  \eea \bea 
G(t_{K}) \to   V  \langle 0 \vert K \vert  
K \rangle_{V} \langle  K \vert K^{\dagger}\vert 0\rangle_{V}  
 \exp( -m_{K}\, t_{K} ) \nonumber \eea
From a study of the time dependence of $G(t,t_{K})$, $G(t)$ and 
$G(t_{K})$ we may extract: 1) the kaon mass 
$m_{K}$; 
2) the two-pion energies on the finite lattice volume, 
$W_{n}$. As shown by M.~L\"uscher in~\cite{ml}, the $W_{n}$ are related 
to the infinite volume phase-shift of the 
two pions, $\delta(k)$,  via the following relations 
\bea &&
W_{n} = 2 \, \sqrt{m_{\pi}^{2} + k^{2} }  \, , \quad \quad 
\frac {\phi (q) + \delta(k)}{\pi} 
 = n \label{phase} \, , \nonumber \\ && n=1,2, \dots \, ,
\quad  q = \frac{k L}{2\pi}  \, , \label{eq:wn}
\eea where $n$ is a non-negative integer~\footnote{ For $n=0$, there
are two solutions: one corresponding to $k=0$ which is spurious,
the other giving the L\"uscher relation between the finite volume
energy and the  scattering length.} and  the function $\phi(q)$ is defined in \cite{ml}.
3)  the operator matrix elements on a finite volume 
$\langle  \pi\pi(n) \vert \hat Q_{i} \vert  K\rangle_{V}$, $\langle  
\pi\pi(n) \vert J \vert 0\rangle_{V}$ and $\langle K \vert K^{\dagger}\vert 
0\rangle_{V} $.
Moreover by a suitable choice of the lattice parameters it is 
possible to match one of the two-pion energies in such a way that 
$m_{K}=W_{n^{*}}$. In practice, since   it will be possible to disentangle only the first few states, 
the matrix elements will be computed with  $n^{*}=0\div 2$.

The fundamental point is that it is possible to relate  the finite-volume Euclidean matrix element 
$\langle  \pi\pi(n^{*}) \vert \hat Q_{i} \vert  K\rangle_{V}$  with 
the absolute value of the physical amplitude  $\langle  \pi\pi \vert \hat 
Q_{i} \vert  K\rangle$:
\bea  && \langle  \pi\pi \vert \hat  Q_{i} \vert  K\rangle = 
\sqrt{{\cal F}} \,   \langle \pi\pi(n^{*}) \vert \hat Q_{i} \vert  K\rangle_{V} 
\, , \nonumber \\  &&
{\cal F}=32 \pi^{2} V^2 \, \frac{\rho_{V}(E) E m_K }{\kappa(E)}\, , 
\label{eq:rela}\eea
where
\bea  \kappa(E)&=&  \sqrt{\frac{E^{2}}{4}-m_{\pi}^{2}} \, , 
\nonumber \\  \rho _{V}(E) &=&\frac{dn}{dE} = \frac{q\phi^\prime(q)
+k\delta^\prime(k)}{4 \pi k^2}E\, . \label{density} \eea
The last expression in
eq.~(\ref{density}) is the one  which one would heuristically
derive from a na\"{\i}ve interpretation of $\rho_V(E)$ as the
density of states, cf. eq.~(\ref{phase}). There are however, some
technical subtleties with such an interpretation which will not be
discussed here. Eq.~(\ref{eq:rela}) holds also when  $m_{K} \neq W_{n}$ and the 
operator carries non-zero energy-momentum~\cite{lmst}.  
By varying the kaon and 
pion masses and momenta, one may  then fit the coefficients of the chiral 
expansion  of $\langle  \pi\pi \vert \hat  Q_{i} \vert  K\rangle$  and 
use these  coefficients to extrapolate to the physical point. This procedure 
is necessary, at  present, since it is not possible  to 
compute the matrix elements at realistic  values of the quark masses and  in the unquenched case. To give an explicit example, the matrix 
elements of $\hat Q_{7,8}$ (for generic meson masses, one pion at rest 
and the other with a given  momentum corresponding to an energy 
$E_{\pi}$)  can be written as~\cite{dlin2001,cirigliano} 
\bea  &&-i {\cal M}_{7,8} (m_{K}, m_{\pi}, E_\pi) =
\gamma^{7,8} \nonumber \\ &&+ \delta^{7,8}_{1}\, \left( \frac{ m_K \, 
(m_\pi+E_\pi)}{2} - m_\pi E_\pi  \right) +\dots \, .\label{eq:eo78} \eea
A fit to the lattice data for ${\cal M}_{7,8}$, extracted from 
suitable correlation functions,  allows  the determination of the 
couplings $\gamma^{7,8}$, $\delta^{7,8}_{1}$, etc. The results of the 
extrapolated amplitudes, if 
chiral logarithms are included, are independent of the cutoff used  in the chiral theory.
\begin{figure}[!t]
\vspace*{-0.5cm}
\mbox{
\hspace*{-0.6cm}
\epsfig{figure=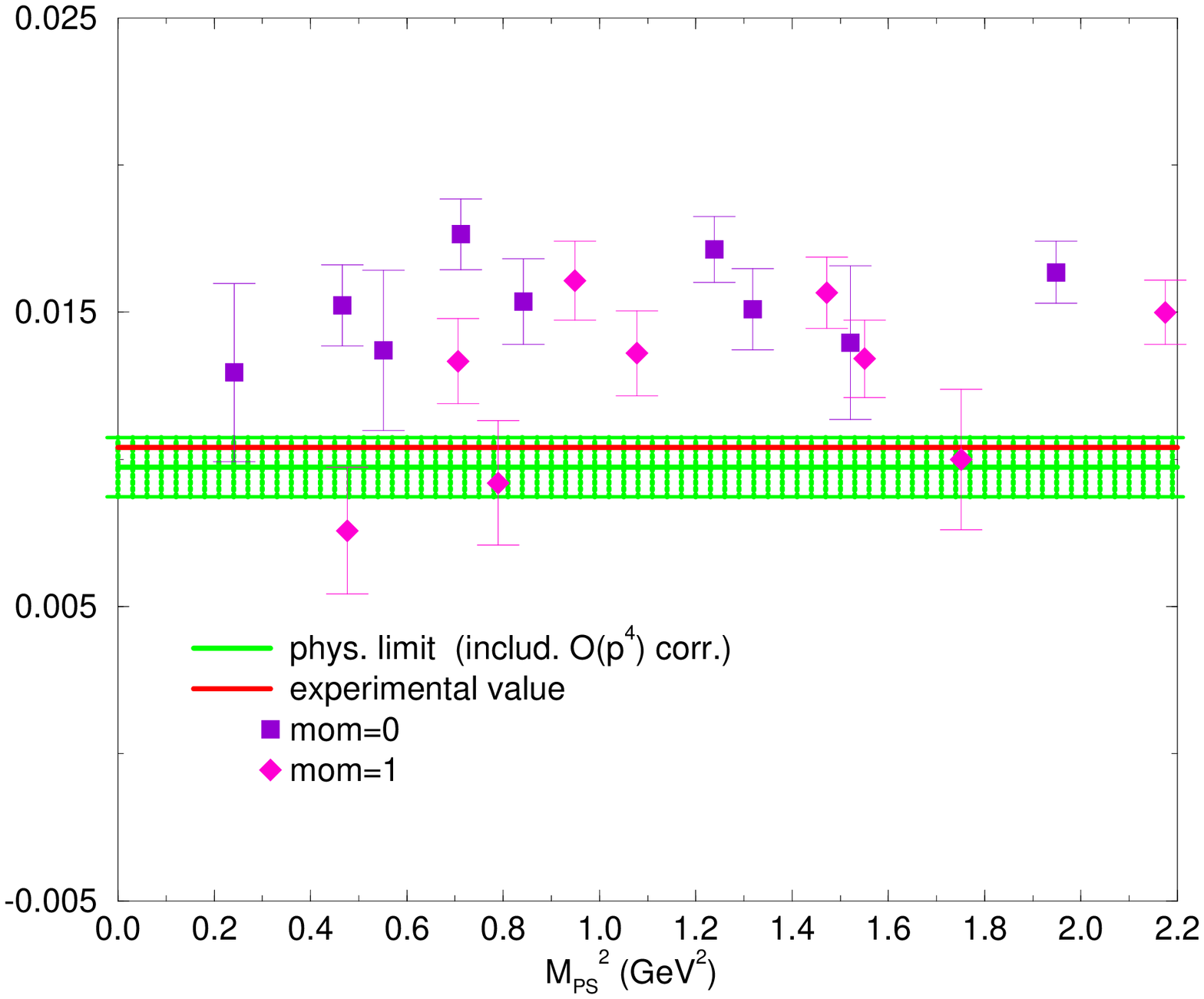,angle=270,width=0.57\linewidth}
\put(0,-13){\epsfig{figure=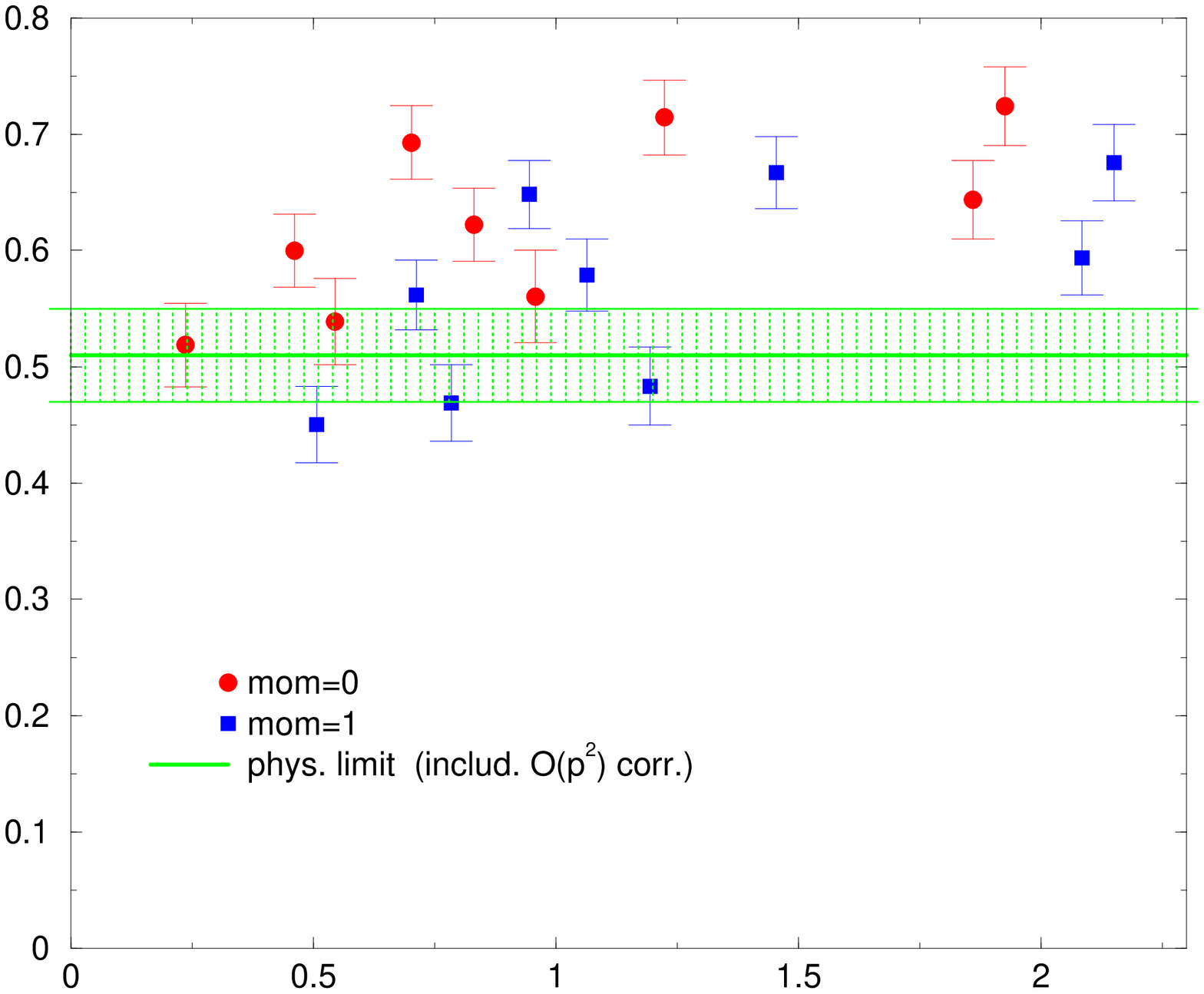,angle=270,width=0.48\linewidth}}}
\vspace{-0.6cm}
\caption{\sl \small a. Chiral behaviour of the matrix element of $Q_{4}$,  ${\cal 
M}_{4} (m_{K}, m_{\pi}, 
 E_\pi)$. The extrapolation to the physical point using only the data 
corresponding to the lightest masses, without the effect of the chiral 
logarithms and using the operators of ref.~\cite{gopa}, is also shown 
as a shadowed band. The experimental result, computed as explained in 
the text,  is also given as a line.  
b. Chiral behaviour of the matrix element of $Q_{8}$,  ${\cal M}_{8} (m_{K}, m_{\pi}, 
 E_\pi)$. The 
extrapolation to the physical point using only the data 
corresponding to the lightest masses, without the effect of the chiral 
logarithms and using the operators of ref.~\cite{cirigliano}, is also shown 
 as a shadowed band.}
\label{o48chiral}\end{figure}\vspace{-0.2cm}
\section{Numerical Results}\vspace{-0.1cm} 
\vspace{-0.1cm} 
\label{sec:numerical}
In this section, the latest numerical results from lattice QCD for $K\to \pi \pi$  and $K\to \pi$  
amplitudes and for $\hat B_{K}$  are  reviewed, together with a 
comparison  with other approaches.  
\subsection*{$\Delta I=3/2$ transitions and $K^{0}$--$\bar K^{0}$ 
mixing}
The SPQ$_{CD}$R Collaboration~\cite{dlin2001,spqr} has presented (quenched) results for 
both $\Delta I=3/2$ and $\Delta I=1/2$ amplitudes, obtained  from a calculation of  $K \to \pi\pi$ matrix elements
following  the strategy of refs.~\cite{lll,lmst}. 
For  $\Delta I=3/2$ transitions,   an extended study of the 
chiral behaviour of the matrix elements of the operators $\hat Q_{4,7,8}$, 
renormalized non-perturbatively, was performed.  In 
figs.~\ref{o48chiral}a  and ~\ref{o48chiral}b,  ${\cal M}_{4} (m_{K}, m_{\pi} 
E_\pi)$   and ${\cal M}_{8} (m_{K}, m_{\pi} 
E_\pi)$   are  shown as a function of the kaon mass. Note that the 
amplitude does  depend on three independent quantities that are let to 
vary, namely $m_{K}$, $m_{\pi}$ and $E_{\pi}$ (one of the two pions is always 
at rest).  In the same figures  the extrapolation to the physical 
point,  performed by  using the chiral expansion of the matrix elements  as in 
eq.~(\ref{eq:eo78}), is given as a band. In  fig.~\ref{o48chiral}a  the 
experimental number, extracted from the $K^{+}  \to \pi^{+} \pi^{0}$ 
decay rate   using the Wilson coefficient of $\hat Q_{4}$ computed at the 
NLO, is also shown. 
The extrapolations are preliminary, since they do not include the 
effects of  (quenched) chiral logarithms that  have not been 
computed yet for the kinematical configurations used in this study.   
For some of the points,  masses and momenta are too large to 
use chiral perturbation theory, and they have not been included in 
the fit.
The  preliminary results in  fig.~\ref{o48chiral}a already  give us an interesting 
physics information. In the chiral limit, and using $SU(3)$ symmetry,  one may relate the $K^{+}  \to 
\pi^{+} \pi^{0}$  amplitude to the $K^{0}$--$\bar K^{0}$ mixing  parameter $\hat B_{K}$~\cite{olddono}.  In this limit, a large 
value for $\hat B_{K}$,  as found in lattice calculations and  unitarity 
triangle analyses~\cite{ciu2001}, $\hat  B_{K}\sim 0.85$, would lead 
to a value of the  $K^{+}  \to  \pi^{+} \pi^{0}$  amplitude   larger than 
the experimental one by $\sim  50\%$.   The extrapolation of the 
results of fig.~\ref{o48chiral}a, together with those for $\hat B_{K}$ 
obtained by the same collaboration~\cite{damir2}  given in 
table~\ref{tab:bks}, demonstrate that chiral corrections to both the  $K^{+}  \to  \pi^{+} \pi^{0}$  amplitude 
and $\hat B_{K}$  can easily reconciliate a large value of $\hat B_{K}$
with the experimental $K^{+}  \to \pi^{+} \pi^{0}$ amplitude.  

It is very instructive to compare the results for $\hat Q_{7}$ and 
$\hat Q_{8}$ 
with those obtained in other approaches, table~\ref{tab:q78}. In the 
case of $\hat Q_{8}$ 
we notice the nice agreement between  lattice results obtained 
from $K \to \pi$ matrix elements using chiral perturbation theory and  
those from the first direct calculation of  $K \to \pi\pi$  
amplitudes.  Similar numbers were obtained by the CP-PACS~\cite{cppacs2001} and RBC~\cite{rbc2001} 
collaborations, which computed $K\to \pi$ matrix elements with DWF. 
These groups presented  the results in different renormalization schemes 
and scales and for this reason the values have not been included in 
the table. There is,  however,  a  very large discrepancy of the lattice 
results with those  obtained with dispersive methods~\cite{dongol} or 
with the $1/N_{c}$  expansion~\cite{derafael},  whose results  would correspond to a huge value of the $B$-parameter,  
$B_{8}=3\div 7$~\footnote{ Other results for $\hat Q_{7}$ and $\hat Q_{8}$  can 
be found in refs.~\cite{bertolini} and \cite{hambye}.}. 
In  order to reproduce the experimental results for 
$\epsilon^{\prime}/\epsilon$ within the Standard Model, such a large 
value of $B_{8}$ implies  a stratosferic value for $B_{6}$. After Kaon 
1999 a new analysis,  performed with spectral function techniques by Bijnens, Gamiz 
and Prades appeared~\cite{Bijnens:2001ps}. In this paper, a value of $\langle 
\hat  Q_{8}\rangle$    much lower than in refs.~\cite{dongol,derafael} was found,
although still about a factor of two larger than  lattice determinations. The very low 
value of $\langle \hat   Q_7\rangle$  found from the lattice $K \to \pi\pi$ 
calculation originates from large cancellations occuring in the 
renormalization of the relevant operator and requires further 
investigation.
\begin{table*}[htb]
\caption{{\sl \small $K \to \pi \pi $ matrix elements in GeV~$^3$, at $\mu=2$~GeV in the 
$\overline{MS}$ scheme, from lattice calculations (first three rows) and other 
approaches.}}
\label{tab:q78}
\newcommand{\m}{\hphantom{$-$}}
\newcommand{\cc}[1]{\multicolumn{1}{c}{#1}}
\renewcommand{\tabcolsep}{2pc} 
\begin{tabular}{@{}llll}
\hline
Quantity& Experiments &Theory \\
\hline
Reference & Method & $\langle \hat   Q_8\rangle$ & $\langle  \hat  Q_7\rangle$  \\
\hline \hline SPQ$_{CD}$R~\cite{spqr} 2001 & $K \to \pi \pi$ &  
$0.53 \pm 0.06$ &  $0.02 \pm 0.01$ \\ 
SPQ$_{CD}$R~\cite{damir2} 2001 & $K \to \pi$ + $\chi$PT &  
$0.49 \pm 0.06$ &  $0.10 \pm 0.03$ \\
APE~\cite{footballteam} 1999 & $K \to \pi$ + $\chi$PT &  
$0.50 \pm 0.10$ &  $0.11 \pm 0.04$ \\
Amherst~\cite{dongol} 1999 & Dispersion relations &  
$2.22 \pm 0.67$ &  $0.16 \pm 0.10$ \\
Marseille~\cite{derafael} 1998  &$1/N_{c}$ & $3.50 \pm 1.10 $ & $0.11 
\pm 0.03$ \\
BGP~\cite {Bijnens:2001ps} 2001  &Spectral functions & $1.2 \pm 0.5 $ & 
$0.26  \pm 0.03$ \\
\hline
\end{tabular}\\[2pt]
\end{table*}

Lattice predictions for $\hat B_{K}$ have been very stable over the 
years, with a central value centered around $0.85$. This is 
essentially also the 
value extracted from Unitarity Triangle 
Analyses~\cite{ciu2001}, thus confirming that lattice QCD can 
predict (and not only postdict) physical quantities. In 
table~\ref{tab:bks} the lattice world average is given, together with 
some of the  latest lattice results. $\hat B_{K}$ from CP-PACS~\cite{cppacsbk} 
(with operators renormalized perturbatively)
and RBC~\cite{rbc2000} (with operators renormalized non-perturbatively)  
has been obtained with DWF. This  formulation of QCD  
should garantee a better control of the chiral behaviour of the 
regularized theory with respect to Wilson-like fermions. The 
results from ref.~\cite{damir2} have been obtained using the 
non-perturbatively improved  Wilson-like action, with a new method based on the 
Ward Identities~\cite{nosub}, which save us from the painful subtractions of the 
wrong chirality operators, which was  necessary in the 
past. Similar strategies have been pursued with 
twisted-mass fermions~\cite{sfm}.  The results with DWF are about $15\%$ below 
the world average and show a marked decrease at small quark masses. 
This could reconciliate the large value of $\hat B_{K}$ at the 
physical kaon mass with the low value of this parameter obtained in 
the chiral limit by  ref.~\cite{derafael1}. Whether the decrease at 
small quark masses is a physical effect, or is due to  lattice artefacts 
(finite volume, residual chiral symmetry breaking etc.) 
remains to be investigated.  New calculations of
 $\langle \bar  K^{0} \vert \hat Q_{i} \vert K^{0} \rangle$  for all 
possible  operators $Q_{i}$ which can mediate $K^{0}$--$\bar K^{0}$ mixing in extensions of the 
Standard Model have  be presented~\cite{damir2}. These results 
are very useful  to put severe constraints on FCNC parameters of SUSY 
models~\cite{footballteam}.
\begin{table*}[htb]
\caption{$B_{K}$ in the 
$\overline{MS}$ scheme at the renormalization scale $\mu=2$~GeV and 
the renormalization group invariant $\hat B_{K}$  from recent lattice calculations. 
The results from SPQ$_{CD}$R, CP-PACS and RBC Collaborations are 
quenched.}
\label{tab:bks}
\newcommand{\m}{\hphantom{$-$}}
\newcommand{\cc}[1]{\multicolumn{1}{c}{#1}}
\begin{tabular}{@{}llll}
\hline
Quantity& Experiments &Theory \\
\hline
Reference & Method & $B_{K}^{\overline{MS}} (2\, \mbox{GeV})$ & $\hat B_{K}$  \\
\hline \hline World Average &  & $0.63\pm 0.03\pm 0.10$ &  
$0.86 \pm 0.06\pm 14$  \\ 
SPQ$_{CD}$R~\cite{damir2} 2001 & with subtractions & $0.71\pm 0.13$ & $0.91 
\pm 0.17$ \\
SPQ$_{CD}$R~\cite{damir2} 2001 & Ward identity method & $0.70\pm 0.10$ & 
$0.90  \pm 0.13$ \\
CP-PACS~\cite{cppacsbk} 2001 & DWF Pert. Ren. &    $0.5746(61)(191)$ &  $0.787 \pm 
0.008$ \\
RBC~\cite{rbc2000} 2000  &  DWF Nonpert. Ren.  &   $0.538 \pm 0.08$ &  $0.737 \pm 0.011$ \\
Ciuchini et al.~\cite{ciu2001} 2001 & Unitarity Triangle &  
$0.70^{+0.23}_{-0.11}$  &  $0.90^{+0.30}_{-0.14}$ \\
\hline
\end{tabular}\\[2pt]
\end{table*}
\begin{figure}[!t]
\vspace*{-0.5cm}
\mbox{{\epsfig{figure=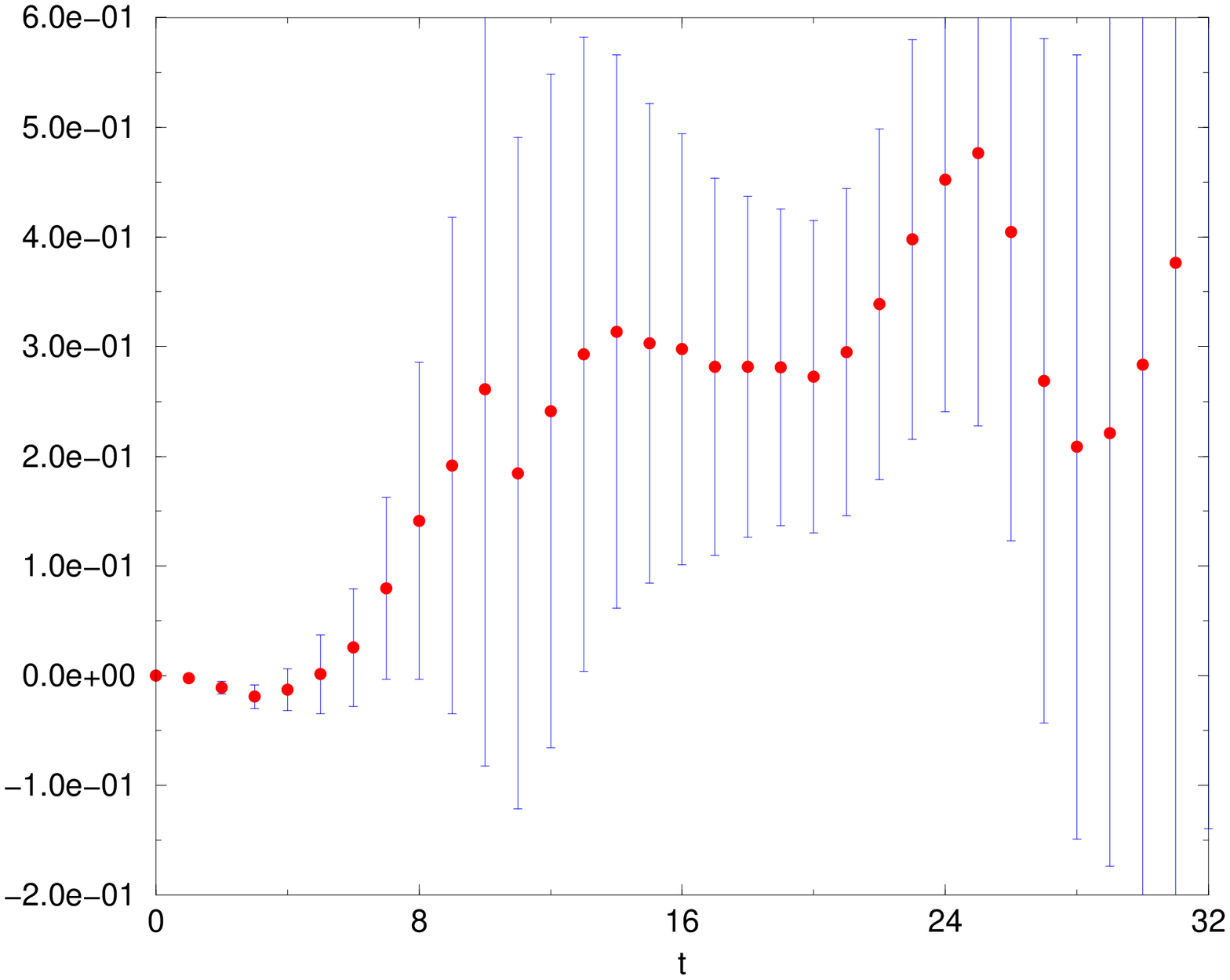,angle=270,width=0.6\linewidth}}}
\vspace{-0.6cm}
\caption{\sl \small 
First signal for the matrix element $\langle \pi \pi 
\vert Q^{-} \vert K \rangle$ at the matching point $m_{K} \sim 
W_{0}$, obtained with 400 configurations  and  $V=24^{3} \times 64$.}\vspace*{-0.4cm}
\label{fig:qm}
\end{figure}\vspace{-0.2cm}
\begin{figure}[htb!]\vspace*{-0.5cm}
\mbox{{\epsfig{figure=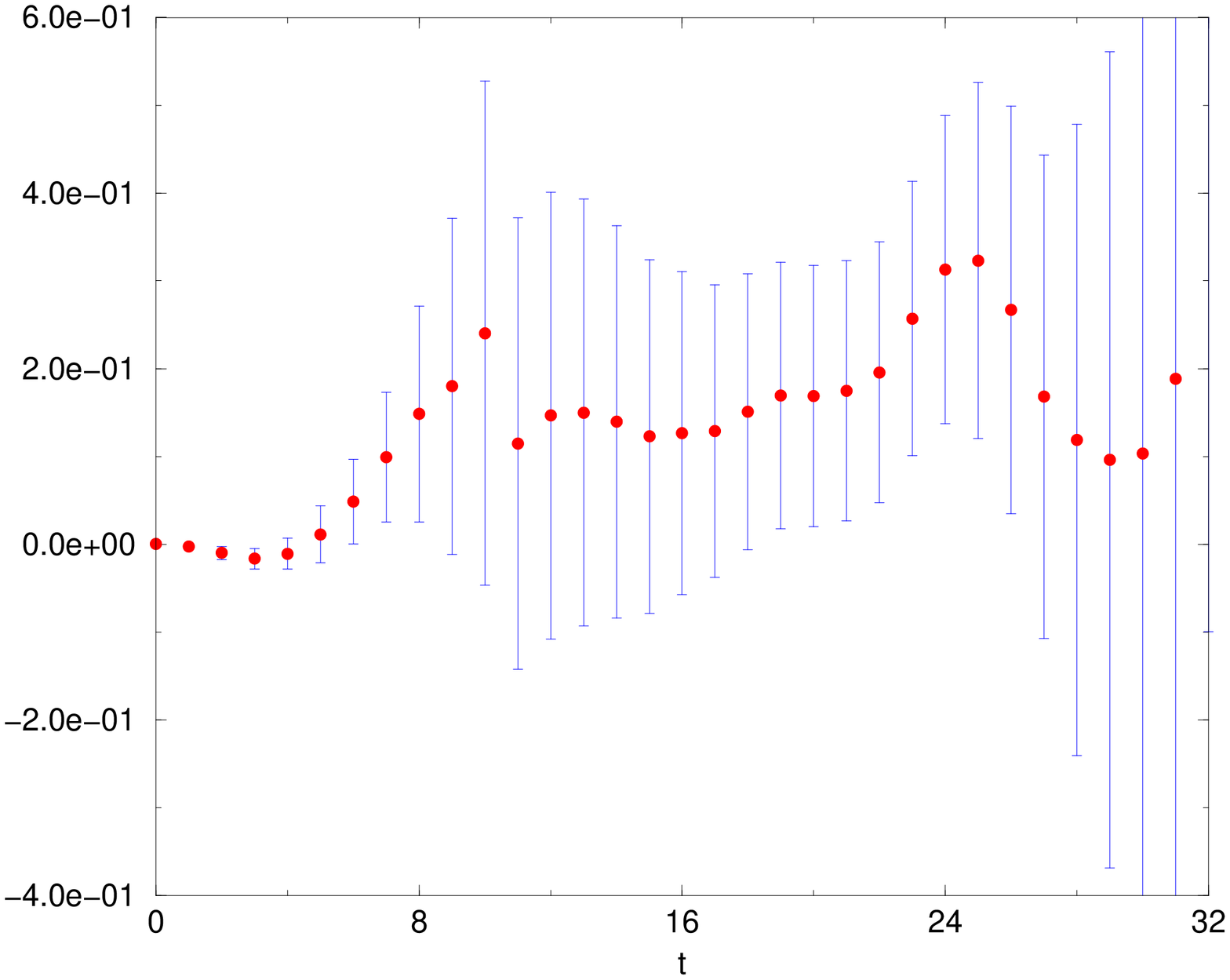,angle=270,width=0.6\linewidth}}}
\vspace{-0.6cm}
\caption{\sl \small 
  First signal for the matrix element $\langle \pi \pi 
\vert Q_{6} \vert K \rangle$ at the matching point $m_{K} \sim 
W_{0}$, obtained with 400 configurations  and  $V=24^{3} \times 64$. }\vspace*{-0.4cm}
\label{fig:q6}
\end{figure}\vspace{-0.2cm}
\begin{figure}[htb!]\vspace*{-0.5cm}
\mbox{{\epsfig{figure=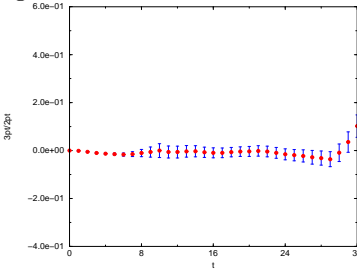,angle=270,width=0.6\linewidth}}}
\vspace{-0.6cm}
\caption{\sl \small 
   $\langle \pi \pi 
\vert \bar s \gamma_{5} d \vert K \rangle$ at the matching point $m_{K} \sim 
W_{0}$, obtained with 400 configurations  and  $V=24^{3} \times 64$. 
At the matching point $\langle \pi \pi 
\vert \bar s \gamma_{5} d \vert K \rangle$ decreases by a huge factor 
and  thus the subtraction for $\langle \pi \pi 
\vert Q_{6} \vert K \rangle$  is very small.}\vspace*{-0.4cm}
\label{fig:qp}
\end{figure}
\subsection*{$\Delta I=1/2$ transitions and  
$\epsilon^{\prime}/\epsilon$}
For  $\Delta I=1/2$ transitions and $\epsilon^{\prime}/\epsilon$,
the subtractions of the power divergencies, necessary to obtain  finite 
matrix elements,   are the major obstacle in lattice calculations. 
These divergencies arise from penguin contractions of the relevant 
operators,
which induce a  mixing with operators of lower dimensions,  
namely $\bar s \sigma^{\mu\nu} G^{A}_{\mu\nu} t^{A} d$, $\bar s 
\sigma^{\mu\nu} \gamma_{5} G^{A}_{\mu\nu} t^{A} d$, $\bar s d$ and 
$\bar s \gamma_{5} d$~\cite{renorm}. Power divergencies and subtractions are 
encountered with both the strategies ($K\to \pi\pi$ or $K \to 
\pi$) adopted to compute the physical  amplitudes. 
For example, with a propagating charm quark,   $Re\, {\cal A}_{0}$ is
computed in terms of the matrix elements of $Q^{\pm}$ ($Q_{1}$ and 
$Q_{2}$)  only. 
In this case, due to the GIM mechanism, the subtraction is implicit  in the difference of 
penguin  diagrams with a charm and an up quark propagating in the 
loop:
\bea Q^{\pm} &=& \bar s \gamma_{\mu} \left(1-\gamma_{5} \right) 
u \, \bar u  \gamma^{\mu} \left(1-\gamma_{5} \right)  d  \\ &\pm& 
 \bar s \gamma_{\mu} \left(1-\gamma_{5} \right) 
d \, \bar u  \gamma^{\mu} \left(1-\gamma_{5} \right) u  - (u \to c)  \, . \nonumber\eea

In the past any attempt to compute $\Delta I=1/2$  $K \to \pi\pi$ matrix elements 
failed because no visible signal was observed after the subtraction of 
the power divergencies~\cite{sonicapri89,marticapri89}. These calculations, performed 
about twelve years ago with very modest computer resources, on small 
lattices and with  little statistics, were abandoned after the 
publication of the Maiani-Testa no-go theorem~\cite{mt}.  

This year, the SPQ$_{CD}$R Collaboration has found the 
first signals for both the matrix elements of $Q^{-}$ and $Q_{6}$, 
figs.~\ref{fig:qm}, \ref{fig:q6} and \ref{fig:qp}.  The 
difference with respect to previous attempts is given by a much larger volume and 
statistics, 
about 400 configurations  with $V=24^{3} \times 64$, compared to a 
few  tens of configurations on lattices with $V=16^{3} \times 32$ (8 
configurations on $V=24^{3} \times 
40$)~\cite{sonicapri89} or 110 configuration of a lattice with $V=16 
\times 12 \times 10 \times 32$ (sic !!)~\cite{marticapri89}. A crucial 
ingredient is also  to work  at the point corresponding to  the matching condition $m_{K} = 
W_{0}$, where $W_{0}$ is the energy of the two pions at rest on the 
finite volume used in the simulation, namely with non degenerate strange 
and light quark masses~\footnote{ 
In this case, as in the case of degenerate quark masses, the 
subtraction of operators of lower 
dimensions is not necessary~\cite{dawson}.}.
 Past calculations were always performed, instead, at the  degenerate point, $m_{K} = m_{\pi}$.   The 
results are very preliminary and in particular the two-pion energy in 
the finite volume for this channel gives a scattering lenght,  
$a_{I=0}$, in disagreement  with expectations  both in value and in the mass dependence.  This is 
to be contrasted with the $\Delta I=3/2$ case, where the analysis of 
the scattering length $a_{I=2}$ is in good agreement with 
expectations~\cite{spqr}. 
Much more work is needed to clarify  these problems  before trying to 
extrapolate the amplitude to the physical point.  

At the matching point $m_{K} = W_{0}$, there are also results 
for  $\langle \pi \pi \vert Q_{6}\vert K \rangle$. In 
this case there is no GIM mechanism at work, and a finite subtraction 
of the matrix element of the pseudoscalar density operator $Q_{P} = 
(m_{s} -m_{d}) \bar s \gamma_{5} d $ must be done (for parity, 
instead, the subtraction of the scalar operator $Q_{S} = (m_{s} + m_{d}) 
\bar s  d $ is not necessary). The  coefficient of the mixing  of $Q_{P}$ with $Q_{6}$ is quadratically divergent~\footnote{ The 
mixing with the chromomagnetic operator is not discussed here. This mixing 
is a small, finite correction  which can be ignored to simplify the  discussion.}
\beq \hat Q_{6} \sim Q_{6} + \frac{C_{P}(\alpha_{s})}{a^{2}} Q_{P} \eeq
If not for the  explicit chiral symmetry breaking of the lattice  
action,  $\langle \pi \pi \vert Q_{P}\vert  K\rangle$  would vanish by the equation of motions  when 
$m_{K} = W_{n^{*}}$. With 
an improved action, $\langle \pi \pi \vert Q_{P} \vert K\rangle$ is 
of ${\cal O}(a^{2})$   and thus a finite subtraction is necessary.  The 
preliminary results shown in fig.~\ref{fig:qp} show that that the value 
of   $\langle \pi \pi \vert \bar s \gamma_{5} d  \vert K\rangle$ drops by a factor of  about 20 with respect 
to  $\langle \pi \pi \vert Q_{6} \vert K\rangle$  
when $m_{K} = W_{0}$ (it is of the same order when the kinematics is not matched)  and thus the subtraction is not very critical. The 
signal itself is very noisy however, see fig.~\ref{fig:q6}, and even with 400 configurations 
the statistical error on  $\langle \pi \pi \vert Q_{6} \vert K\rangle$  is about  $50\, \%$. Thus 
 a rather large statistics will be  necessary to obtain a relatively accurate result. 

\begin{figure}[!t]
\vspace*{-0.5cm}
\mbox{{\epsfig{figure=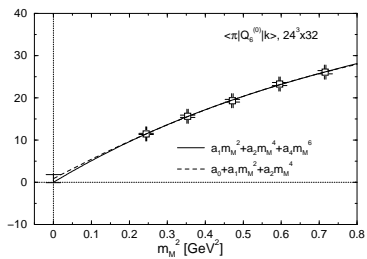, width=0.6\linewidth}}}
\vspace{-0.6cm}
\caption{\sl \small 
Chiral behaviour of the  matrix element $\langle 
\pi \vert Q_{6} \vert K \rangle$ from the CP-PACS 
Collaboration. The curves represent different fits used to extrapolate 
to the chiral limit.}\vspace*{-0.4cm}
\label{fig:cppacschiral}
\end{figure}

$K \to \pi \pi$ amplitudes in the chiral limit can also be 
extracted from the calculation of $K\to \pi$ matrix elements, and indeed  
this has been the most popular method in  lattice 
calculations of $\Delta I =3/2$ transitions.  The main advantage of  this approach is that the 
three-point correlators, necessary for the extraction of the matrix 
elements, are less noisy than the four-point correlators used in the  $K\to 
\pi\pi$ case. Moreover, for  $K\to \pi$ matrix elements,  finite volume corrections  
are exponentially small as $L \to \infty$.   The main disadvantage is 
that the $K \to \pi\pi$  amplitudes are obtained, using soft pion 
theorems,  in the chiral limit only  and then the 
effect of FSI is definitively lost.  If these are important for  $\Delta 
I=1/2$ channels there is no hope, then, to recover the physical 
amplitudes. 
In order to make the $K \to \pi$ matrix element finite  at least a subtraction 
is necessary,  even in the absence of explicit chiral symmetry breaking in the 
action.    With Wilson-like  Fermions, for which chiral symmetry is explicitly broken, 
the  number of subtractions  makes this approach unpracticable as 
also demonstrated  by the failure  of  past  
attempts~\cite{sonicapri89,marticapri89}. The  
method has acquired a renewed popularity, instead,  with 
recent formulations of the lattice theory, DWF or 
overlap Fermions,  for which chiral symmetry breaking is absent 
or   strongly reduced in practice.     In this case, and under the 
hypothesis that chiral symmetry is under control, a finite amplitude 
can be obtained  by subtracting the scalar density amplitude with a 
suitable coefficient $C_{i}$ which depends on the operator at hand
\beq \langle \pi \vert Q_{i}^{sub} \vert K \rangle = \langle \pi \vert Q_{i} \vert K \rangle  - C_{i} \langle \pi \vert Q_{S}\vert K \rangle \, .
\label{eq:subkpi}  \eeq
The power divergent coefficients $C_{i}$ cannot be computed 
perturbatively.  Following  ref.~\cite{wise}, when  chiral symmetry is 
preserved,  the $C_{i}$ can be determined  by  the condition that the $K \to 0$   matrix element 
of the subtracted operator vanishes
\beq \langle 0 \vert Q_{i}  - C_{i} Q_{P} \vert K \rangle =0 \, . \eeq
The coefficients $C_{i}$ have been  obtained either  using  non degenerate 
quarks, $m_{s} \neq m_{d}$, by the RBC Collaboration,  or from the 
derivatives  of the 2-point correlation function with respect to the quark mass, 
by  the CP-PACS Collaboration.
\begin{figure}[!t]
\vspace*{-0.5cm}
\mbox{{\epsfig{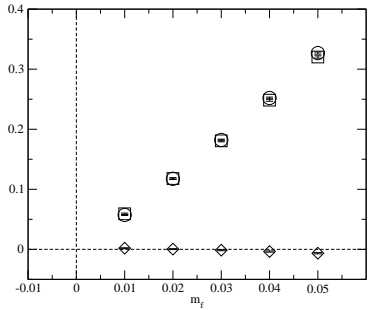}}}
\vspace{-0.6cm}
\caption{\sl \small 
 $\langle  \pi 
\vert Q \vert K \rangle$ for the unsubtracted operator,  $Q_{6}$ 
(squares),  the subtraction, $C_{6} Q_{S}$ (Circles), and the total, 
$Q_{i}^{sub}$ see eq.~(\ref{eq:subkpi}), as a function  of the quark mass 
$m_{f}$. The results are from the RBC Collaboration.}\vspace*{-0.4cm}
\label{fig:subrbc}
\end{figure}
Numerically, the  matrix element of the subtracted operator is much smaller than the 
unsubtracted one, as shown by  fig.~\ref{fig:subrbc}, taken from the 
recent work of the RBC collaboration~\cite{rbc2001}. This implies that any systematic uncertainty which 
enters in the subtraction procedure can have huge effects in the 
determination  of the physical amplitudes. After the subtraction, the 
chiral dependence of the matrix element has to be fitted in order to 
extrapolate it to the chiral limit. Both groups have included the 
logarithmic corrections which arise in quenched $\chi$PT in the fit 
and, in some cases, polynomial corrections of higher order in 
$m^{2}_{\pi}$. The chiral behaviour observed by  CP-PACS  is very 
satisfactory, as shown by fig.~\ref{fig:cppacschiral}, less good in the 
case of the RBC Collaboration. The difference may be due to the fact 
that a different gluon action is used in the two cases,  corresponding 
to smaller chirally breaking effects for CP-PACS. This can be monitored by 
measuring the so called ``residual mass'' which should be zero for 
perfect chirality and  is about a factor of ten smaller for CP-PCAS 
than for RBC. For a more extended discussion on this important point 
the reader can refer to~\cite{cppacs2001,rbc2001}.  
Let me also mention that the overall 
renormalization constants have been computed perturbatively by 
CP-PACS~\cite{aoki} and non-perturbatively by 
RBC~\cite{rbcnp}.
My compendium  of the physics results obtained by the two groups is given 
in table~\ref{tab:rbcpacs}.
A few observations are necessary at this point. First of all,  $Re {\cal 
A}_{0}$, and consequently $Re {\cal 
A}_{0}/Re {\cal A}_{2}$, from the RBC  Collaboration are in good agreement with 
experimental values, contrary to CP-PACS which finds $Re {\cal 
A}_{0}$ smaller   by about a factor of two than the experimental number. 
Since the two groups use the same lattice formulation of the theory 
and differ only by, hopefully, marginal details, the reason of this 
difference should be clarified.  In both cases, however,  $\epsilon^{\prime} / 
\epsilon$ is in total disagreement with the data. The main reason is 
that the value of the matrix element  of $Q_{6}$ is much smaller than 
what would be necessary to reproduce the experimental value  (it corresponds 
approximatively to $B_{6} = 0.3\div 
0.4$).
Let me list a number of sources of systematic errors which may explain 
these embarassing results:

1) \underline{Chirality} By working with DWF at a finite fifth 
dimension, $N_{5} =16$,  
a residual chiral symmetry  breaking is present in the theory. The 
amount of residual symmetry breaking is parametrized by a mass scale 
denoted by $m_{res}\sim 0.2 \div 2.0$~MeV. In the 
presence of explicit chiral symmetry breaking, the coefficients 
$C_{i}$ determined from $K \to 0$  differs from the correct one and 
this may induce an  error of ${\cal O}(m_{res} a^{-2}) \sim 
(200 \, 
\mbox{MeV})^{3}$ on matrix elements which are of the order of 
$\Lambda_{QCD}^{3} \sim (300  \, \mbox{MeV})^{3}$.   Both groups 
claim to have this point fully under control~\cite{cppacs2001,rbc2001}. 
A calculation at a larger value of $N_{5}$, with all the other 
parameters unchanged, would be very useful to clarify the situation. 

2) \underline{Matching below the charm mass} Both groups have
 so far presented results at a renormalization 
scale just  below the charm quark mass. The  matching of the effective
 theory is rather problematic at such low scales.

3) \underline{Extra Quenched Chiral Logarithms} As 
shown by fig.~\ref{fig:subrbc}, the subtraction of the power divergencies is very critical. 
Besides the effects discussed in 1), 
 another delicate problem   has been  recently rised by 
Golterman and Pallante~\cite{goltepallalast}. In the quenched theory  the 
$(8,1)$ operators, such as $Q^-$ and $Q_{6}$, do not belong anymore to 
irreducible representations of the chiral group and this gives rise to spurious chiral logarithms. These logarithms affect the subtraction 
procedure and   have not been taken into account in  the analyses of RBC 
and CP-PACS.  A preliminary numerical study to try   to get rid of 
these effect has been discussed at this conference~\cite{sharpeboh}   

4) \underline{Higher order chiral corrections and FSI} Higher 
order chiral contributions, among which FSI have also to be accounted, 
can produce  large variations  of the matrix elements  between the 
chiral limit and the physical point.  In simulations where only  $K \to \pi$ 
matrix elements are computed,  $K \to \pi\pi$  amplitudes can only be 
obtained at lowest order in $\chi$PT where these physical effects are 
missing.

5) \underline{New Physics} As noticed by Murayama and 
Masiero~\cite{murayama}, it 
is possible to produce large effects for $\epsilon^{\prime} /\epsilon$ 
in SUSY without violating other bounds coming from FCNC. If the 
lattice results are correct, this is an open possibility. Before 
invoking new physics,   at least the result for $Re{\cal 
A}_{0}$ should be established with more confidence. Given that the two 
groups still don't agree on this quantity, more work is needed.
\section{Conclusion and outlook}
\vspace{-0.1cm} 
In this review, because of space limitations, I discussed only a 
subset of the results for light-quark matrix elements presented at 
Lattice 2001. Although the focus has been on kaon weak matrix 
elements, the general discussion relative to renormalization, 
improvement and calculation of physical amplitudes applies to many 
other cases as well. 
A renewed activity in lattice calculations of kaon decays and mixing  
has developed in the last two years.
For $\Delta I=3/2$ transitions, $K^{+} \to \pi^{+} \pi^{0}$ and 
electroweak penguin amplitudes, new and more  precise results have 
been obtained. In this case, removal of lattice artefacts, by 
extrapolation to the continuum and/or improvement, and unquenched 
calculations are around the corner.
For $\Delta I=1/2$ transitions,  it has been shown that direct computations 
of  $K \to \pi\pi$ amplitudes, including FSI,  is 
possible~\cite{lll,lmst}, although 
the pratical implementation with sufficient accuracy will require more 
time.  First results, obtained by  using $K \to \pi$ lattice matrix elements 
computed with DWF and soft-pion theorems,  show a striking 
disagreement with the experiments for $\epsilon^{\prime}/\epsilon$.  
For $Re {\cal A}_{0}$,  there is  a factor of two between the values 
found by the CP-PACS~\cite{cppacs2001}  and 
RBC~\cite{rbc2001} Collaborations. More work is needed in this case to clarify all 
these points. Many related physical quantities, like the strong 
interaction  phase-shifts for $\pi \pi$ scattering, the 
chromomagnetic operator matrix elements,  semileptonic $K \to \pi\pi 
l \nu_{l}$ amplitudes and the scalar semileptonic form factor will also be 
obtained as a byproduct from  lattice investigations of non-leptonic kaon 
decays. In particular, the study of the $K \to \pi\pi 
l \nu_{l}$ amplitude is very useful since it contains all the 
ingredients relative to FSI without having the extra difficulties due 
to the power divergences of the non-leptonic amplitudes,  and the 
corresponding necessary subtractions. 
\section*{Acknowledgements}
I thank  particularly D.~Becirevic, M.~Golterman R.~Gupta, D.~Lin, 
V.~Lubicz, R.~Mawhinney,
J.~Noaki, E.~Pallante, M.~Papinutto and  S.~Sharpe for  fruitful and   constructive   discussions. 
I am also particularly indebted to all the members of the SPQ$_{CD}$R 
Collaboration.  This work was supported by European Union grant HTRN-CT-2000-00145.

\end{document}